\documentclass[]{article}

\usepackage[affil-it]{authblk}
\usepackage{a4wide}
\usepackage{amsmath}
\usepackage{multirow}
\usepackage{booktabs}
\usepackage[ruled,vlined]{algorithm2e}
\usepackage{array}
\usepackage{color}
\usepackage[normalem]{ulem}
\usepackage{hyperref}
\usepackage{amssymb}
\usepackage{graphicx}
\usepackage{amsfonts}
\usepackage[english]{babel}
\usepackage[sort&compress,square,comma,authoryear]{natbib}

\begin{document}
\bibliographystyle{apalike}

\title{Implementation of 3D degridding algorithm on the NVIDIA GPUs using CUDA}

\author[1]{Karel Ad{\'a}mek}
\author[2,3]{Peter Wortmann}
\author[2]{Bojan Nikolic}
\author[3]{Ben Mort}
\author[4]{Wesley~Armour}
\affil[1]{Faculty of Information Technology, Czech Technical University, Th\'{a}kurova 9, 160 00, Prague, Czech Republic}
\affil[2]{Astrophysics Group, Cavendish Laboratory, JJ Thomson Avenue, Cambridge CB3 0HE, United Kingdom}
\affil[3]{SKA Organisation, Jodrell Bank, Lower Withington, Macclesfield, Cheshire, SK11 9FT, United Kingdom}
\affil[4]{Oxford e-Research Centre, Department of Engineering Sciences, University of Oxford, 7 Keble road, OX1 3QG, Oxford, United Kingdom}

\maketitle

\begin{abstract}
Practical aperture synthesis imaging algorithms work by iterating between estimating the sky brightness distribution and a comparison of a prediction based on this estimate with the measured data ("visibilities"). Accuracy in the latter step is crucial but is made difficult by irregular and non-planar sampling of data by the telescope. In this work we present a GPU implementation of 3d de-gridding which accurately deals with these two difficulties and is designed for distributed operation. We address the load balancing issues caused by large variation in visibilities that need to be computed. Using CUDA and NVidia GPUs we measure performance up to 1.2 billion visibilities per second.
\end{abstract}

\section{Introduction}
Degridding is essential to deconvolution methods like CLEAN \citep{1983AJ.....88..688S, 2008ISTSP...2..793C} which iteratively recovers the sky brightness distribution by comparing measured data called visibilities with model of the sky. In this process the sky brightness distribution is first calculated from the irregularly sampled visibilities. The brightest components of the sky brightness distribution are added into the sky model which represent an approximation to sky brightness distribution. When the sky model is updated, regularly sampled visibilities of the model must be degridded back to where the measured visibilities have been sampled and then subtracted from the data. The process is repeated on resulting residuals until convergence. The accuracy in the degridding step is crucial for accurate image of the sky.

Degridding is part of many imaging pipelines \citep{2014MNRAS.444..606O, 2013A&A...553A.105T}. The GPU implementation of the image domain degridding algorithm was published by \citep{7967145}. In this work, we have examined the GPU implementation of a deconvolution\footnote{The source code is available at \url{https://github.com/KAdamek/GPU_degridding}} of a list of visibilities with the pre-computed grid data which is a part of the improved w-stacking \citep{2019PhDT........15Y}. 


\section{Implementation}
We have implemented the degridding using CUDA for NVIDIA GPUs. The degridding algorithm is a sum of weighted regularly sampled visibilities. In case of 3d degridding we assume that the regularly sampled visibilities with coordinates $u$, $v$, and $w$ are in the form of a stack of subgrids. For the implementations we have further assumed that on the input there is:
\begin{itemize}
\item{precomputed oversampled gridding convolution function (GCF) for $uv$ plane $G_{uv}$, with support $8\times8$;}
\item{precomputed oversampled GCF for $w$ planes $G_{w}$, with support $4$;}
\item{$S$ precomputed subgrid stack of size $512\times512\times4$ where each subgrid contains $N_\mathrm{vis,S}$ visibilities which need to be deconvolved; and}
\item{a list of $N_\mathrm{vis,S}$ visibility coordinates $[u,v,w]$ for each subgrid.}
\end{itemize}
To test the performance, we have assumed that the visibilities which need to be deconvolved form a line with a constant step $\mathrm{d}u$, $\mathrm{d}v$, and $\mathrm{d}w$. However, the code does not rely on this assumption and work with any distribution of visibilities.

For each visibility $i$ from a subgrid $S$ we need to calculate two sets of coordinates. One set of coordinates $g=[x,y,z]$ refers to the position of a subgrid point where the deconvolution starts. The second set of coordinates is for the location of correct weights from the oversampled gridding convolution function GCF $G_{uv}$ and $G_{w}$. These coordinates are calculated within the threadblock. 

In the optimised version of the deconvolution (D1) a single threadblock of 32 threads processes $N_b$ visibilities which are adjustable. This offers some data reuse as a single grid point may contribute to a multiple visibilities. The data reuse of the weights is harder due to oversampling of the GCF. The threads first cooperate on the calculation of all coordinates for assigned visibilities. These values are then stored in the shared memory and used by all threads in the deconvolution part. 

The deconvolution processes in two phases. First, each thread accumulates its complex visibility along subgrids $y$-axis ($v$-axis). This allows threads to have contiguous access to memory for both subgrid point and GCF weights. In the second phase, these partial results are pooled using parallel reduction to get deconvolved visibility. As all these operations are confined to a single CUDA warp no synchronization is required. 

The visibilities are stored in the shared memory and written out in bulk in a contiguous manner to the device memory at the end of the threadblock execution.

The number of threadblocks in the $x$-direction of the CUDA grid is given by the maximum number of visibilities per subgrid from all active subgrids
\begin{equation}
N_\mathrm{th}=\max_S{\left(N_\mathrm{vis,S}/N_b\right)}\,.
\end{equation}
The number of threadblocks in $y$-direction is equal to the number of subgrids $S$. This distribution of the work between threadblock may pose a load-balancing issue as not all subgrids may have the same amounts of visibilities. This may be a problem for SKA as some of the subgrids may have significantly more visibilities than others. 

In addition to the deconvolution described above, we have also implemented different variations of the algorithm. Alternatively, the optimised GPU kernel could be launched with 64 threads and the deconvolution is first reduced in $z$-direction ($w$-direction) before all threads perform parallel reduction (kernel D4). Since this involves multiple warps it requires synchronization. 

Lastly, we have implemented a Basic version (D2) of the GPU kernel where each threadblock processes only one visibility (D2). This GPU kernel is also implemented using dynamic parallelism (D3). Dynamic parallelism allows us to launch GPU kernels from inside of another GPU kernel. This means that each thread launches a separate CUDA grid with different dimensions for each subgrid mitigating load-balancing to a degree.


\begin{figure}
\centering
\includegraphics[width=1.0\textwidth]{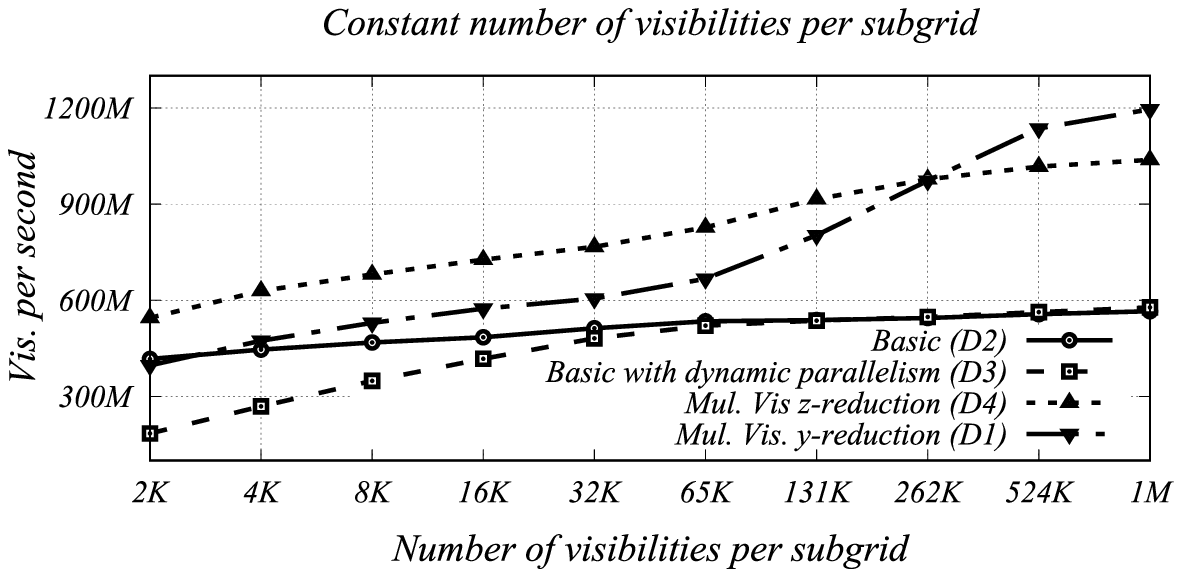}
\caption{Number of visibilities processed per second for subgrids with constant number of visibilities per subgrids. \label{ex_fig1}}
\end{figure}


\begin{figure}
\centering
\includegraphics[width=1.0\textwidth]{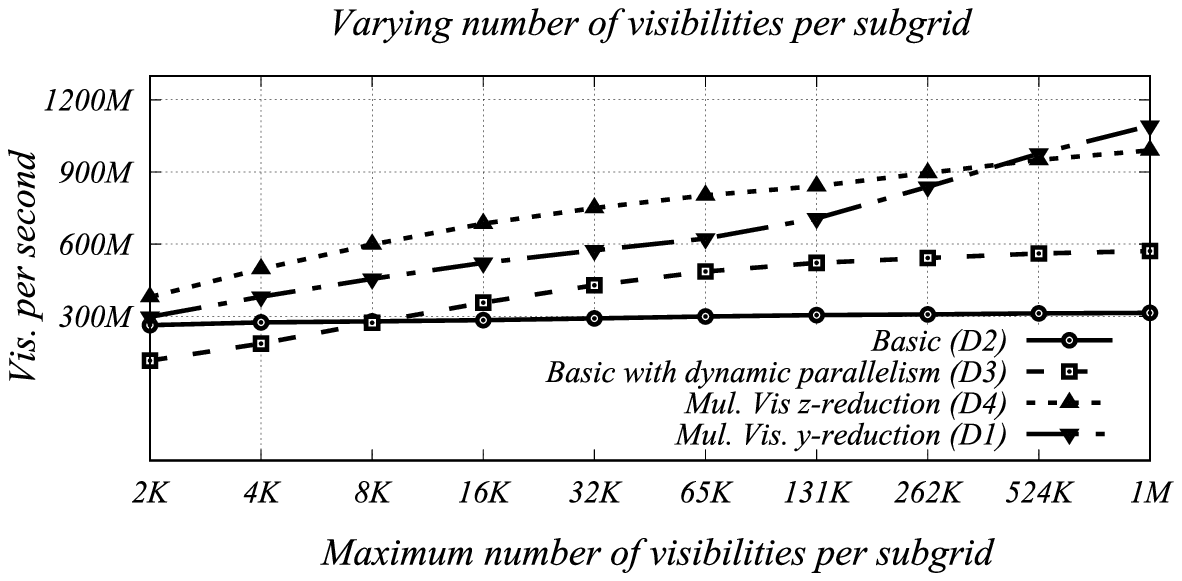}
\caption{Number of visibilities processed per second for subgrids with varying number of visibilities per subgrids. \label{ex_fig2}}
\end{figure}

\section{Results}
To assess the performance of the deconvolution GPU kernel we have used two scenarios. In the first scenario, the number of visibilities per subgrid is the same for all subgrids. This is shown in Figure~\ref{ex_fig1}. In the second scenario, shown in Figure~\ref{ex_fig2} the number of visibilities per subgrid is increasing with every subgrid up to the maximum number of visibilities in the last subgrid.

The results show that the choice of the optimal GPU kernel depends on the expected number of visibilities per subgrid. For a lower number of visibilities, below 262 thousand, reduction through $z$-direction (D4) is better. For a higher number of visibilities, the better performing kernel is kernel D1. The results also show the advantage of dynamic parallelism for basic GPU kernel which was not repeated when applied to optimised kernels D1 or D4. The optimal number of visibilities processed per threadblock $N_b$ depends on the step $\mathrm{d}u$, $\mathrm{d}v$, and $\mathrm{d}w$. The differences are however marginal and single value of $N_b=32$ is sufficient. For the number of visibilities per subgrid below 16 thousand, it is substantially better to process only one visibility per threadblock.

\section*{Acknowledgements} 
This work has received support from STFC Grant (ST/T000570/1). The authors acknowledge the support of the OP VVV MEYS funded project CZ.02.1.01/0.0/0.0/16\_019/0000765 "Research Center for Informatics". 


\end{document}